\documentclass[aps,pra,floatfix,twocolumn,tightenlines]{revtex4}
\usepackage{amsmath}
\usepackage{amsfonts}
\usepackage{latexsym}
\usepackage{graphicx}
\usepackage{color}
\usepackage[bookmarks=true,colorlinks=true,urlcolor=blue]{hyperref}
\usepackage{multirow}

\DeclareMathOperator{\Tr}{tr}
\newcommand*{\cW}{\mathcal{W}}
\newcommand{\SH}{\mathbf{H}}
\newcommand{\SV}{\mathbf{V}}
\newcommand{\SD}{\mathbf{+}}
\newcommand{\SA}{\mathbf{\!-\!}}
\newcommand{\SL}{\mathbf{L}}
\newcommand{\SR}{\mathbf{R}}

\newcommand{\ket}[1]{|#1\rangle}

\newcommand{\proj}[1]{|#1\rangle\langle #1|}
\newcommand{\EV}[1]{\langle #1\rangle}

\begin{document}

\title{Experimental bound entanglement in a four-photon state}
\author{Jonathan Lavoie}
\affiliation{Institute for Quantum Computing and Department of Physics \&
Astronomy, University of Waterloo, Waterloo, Ontario, Canada, N2L 3G1}
\author{Rainer Kaltenbaek}
\affiliation{Institute for Quantum Computing and Department of Physics \&
Astronomy, University of Waterloo, Waterloo, Ontario, Canada, N2L 3G1}
\author{Marco Piani}
\affiliation{Institute for Quantum Computing and Department of Physics \&
Astronomy, University of Waterloo, Waterloo, Ontario, Canada, N2L 3G1}
\author{Kevin J. Resch}
\email{kresch@iqc.ca}
\affiliation{Institute for Quantum Computing and Department of Physics \&
Astronomy, University of Waterloo, Waterloo, Ontario, Canada, N2L 3G1}

\begin{abstract}

Entanglement~\cite{Schroedinger1935a, HHHH09} enables powerful
new quantum technologies~\cite{ekert1991quantum, bennett1993teleporting, Mattle1996a, Bouwmeester1997a, Jennewein2000a,Nielsen2000a}, but in real-world
implementations, entangled states are often subject to
decoherence and preparation errors. Entanglement
distillation~\cite{Bennett1996b,bennett1996mixed} can often
counteract these effects by converting imperfectly entangled
states into a smaller number of maximally entangled states.
States that are entangled but cannot be distilled are called
\textit{bound entangled}~\cite{horodecki1998mixed}. Bound entanglement is central to many exciting theoretical results in quantum information processing~\cite{Horodecki2005a, Horodecki2008a,Smith2008a}, but has thus far not been experimentally realized. A recent claim for experimental bound entanglement is not supported by their data~\cite{amselem2009experimental}.
Here, we consider a family of four-qubit Smolin states~\cite{smolin2001four}, focusing on a regime where the bound entanglement is experimentally robust. We encode the state into the polarization of four photons and show that our state exhibits both entanglement and undistillability, the two defining properties of bound entanglement. We then use our state to implement entanglement unlocking, a key feature of Smolin states~\cite{smolin2001four}.

\end{abstract}

\maketitle

Bound entangled states are important for
several reasons. First, they represent irreversibility in
entanglement manipulation: they require the consumption of pure
entanglement to be created via local operations and
classical communication (LOCC), but no pure entanglement can be
distilled from them via
LOCC~\cite{horodecki1998mixed,yang2005irreversibility,piani2009relative,brandao2010generalization}.
Second, they constitute a challenge to develop better
entanglement criteria, as there is no standard efficient way to
detect their entanglement~\cite{horodecki1997separability,HHHH09}. Third, they are central to recent breakthroughs regarding
channel capacities~\cite{Smith2008a}. Fourth, despite not being
distillable, they still constitute a resource for quantum
teleportation~\cite{masanes2006all}, quantum
cryptography~\cite{Horodecki2005a,Horodecki2008a}, and channel
discrimination~\cite{piani2009all}. Thus, bound
entanglement is crucial for developing a more complete picture
of the role of entanglement in quantum information.

One of the most elegant and striking examples of bound
entanglement is the four-party Smolin
state~\cite{smolin2001four},
\begin{equation}
\rho_S=\frac{1}{4}\sum_{\mu=0}^3|\Psi^\mu\rangle\langle\Psi^\mu|_{AB}\otimes|\Psi^\mu\rangle\langle\Psi^\mu|_{CD},
\label{eqn:smolin}
\end{equation}
where the subscripts label the parties, and $|\Psi^\mu\rangle$
are the two-qubit Bell states. One may understand the Smolin state in the
following way: $A$ and $B$ share one of four possible Bell
states, and $C$ and $D$ share the same state, but each Bell
state is equally likely and unknown. The Smolin state is
entangled in the sense that it does not admit a fully-separable
decomposition of the form $\sum_kp_k\rho^k_A\otimes\eta^k_B\otimes \tau^k_C\otimes \xi^k_D$, with $p_k$ probabilities and $\rho^k,\eta^k,\tau^k,\xi^k$ states. On the other hand,
no pure entanglement can be distilled, neither in the form of
bipartite nor multipartite pure entangled
states~\cite{smolin2001four}. Thus, the Smolin state is, by
definition, bound entangled. Despite its undistillability, the entanglement in the Smolin
state is unlockable in the sense that a joint measurement
between any two parties can enable a pure maximally entangled
state between the other two~\cite{smolin2001four}.

It is evident from equation~(\ref{eqn:smolin}) that the Smolin
state is separable (or unentangled) in the $(AB):(CD)$
bipartite cut since it can be written in a biseparable form
$\rho=\sum_k p_k\rho^k_{AB}\otimes\tau^k_{CD}$. As the state is
symmetric with respect to the exchange of any two
parties~\cite{smolin2001four,augusiak2006bound}, it is separable with respect to all three two-two bipartite cuts $(AB):(CD)$, $(AC):(BD)$, and $(AD):(BC)$. Following the
arguments in~\cite{smolin2001four}, one concludes that no
entanglement can be distilled between any two parties, and this
excludes also the distillation of three- and four-partite
entanglement. Similarly, to prove that an experimentally
prepared Smolin state is undistillable, it is sufficient to show that all
eigenvalues remain positive under partial transposition (PPT) across all two-two bipartite cuts~\cite{horodecki1998mixed}.

A recent work reported the production of a pseudo-bound
entangled state in liquid-state NMR~\cite{Kampermann09nmr}.
They demonstrate sufficient control over their system to
implement the transformations that lead to bound entanglement.
Yet, they would need to start in a highly pure, rather than a
highly mixed state, to generate bound, as opposed to
pseudo-bound, entanglement. Optical systems on the other hand
can produce highly pure states. In
Ref~\cite{amselem2009experimental}, the authors claimed an
optical demonstration of a Smolin bound-entangled state. Their
data showed that their state was entangled. They reconstructed
their state using quantum state tomography and applied the PPT
test for distillability and argued that the eigenvalues are all
positive or consistent with zero. However, their results show five negative
eigenvalues over all three partial transpositions with a minimum
of $-0.02\pm0.02$. Since the central point in the PPT test is
that the minimum eigenvalue has to be \textit{non-negative}, their claim
of demonstrating bound entanglement is not supported by their
data.

\begin{figure}
\begin{center}
\includegraphics[width=0.85 \columnwidth]{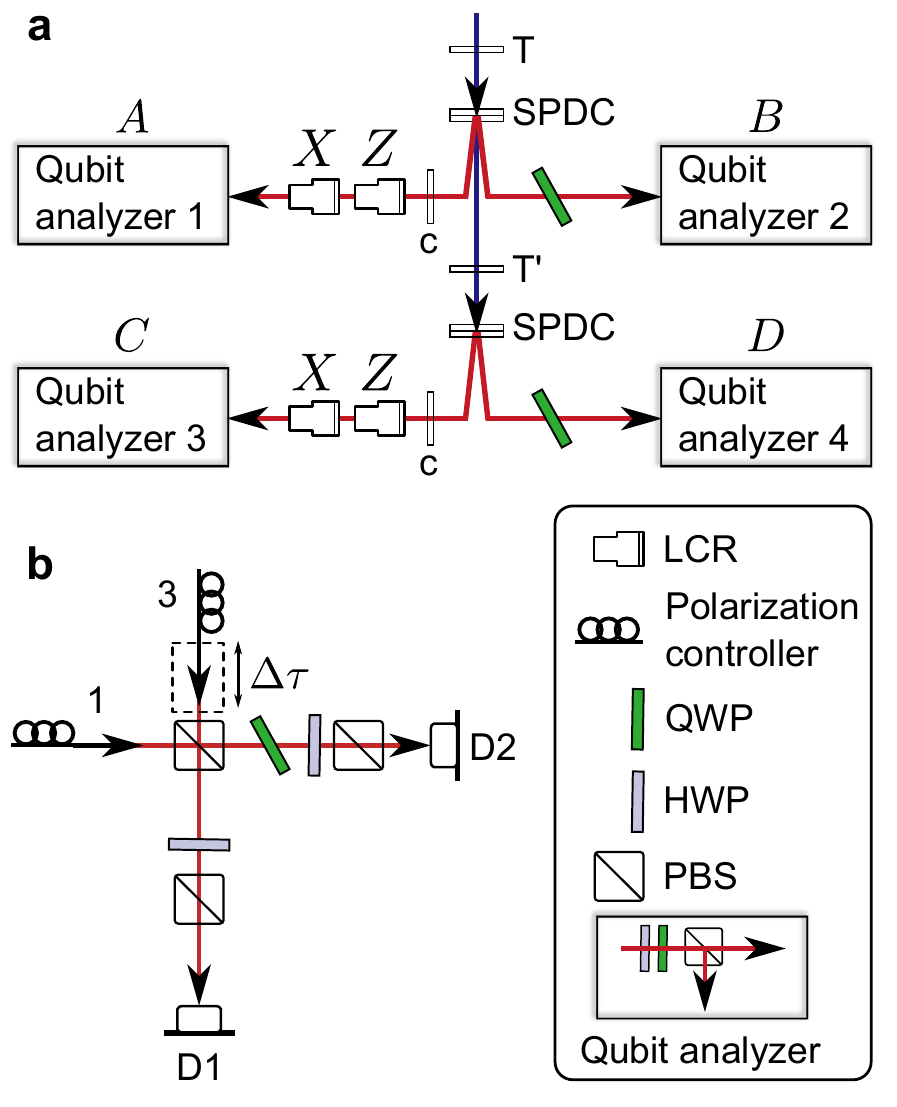}
\caption{\textbf{Experimental setup to generate four-photon Smolin states.} \textbf{a,} We generate a family of Smolin states
by randomly applying unitaries, using two pairs of liquid-crystal variable phase retarders (LCRs), to the initial
$\ket{\phi^+}_{AB}\otimes\ket{\phi^+}_{CD}$ state produced by
two down-conversion sources. For each party $A$, $B$, $C$ and
$D$, the polarization is analyzed with a combination of a
half-wave plate (HWP), quarter-wave plate (QWP) and polarizing
beam-splitter (PBS). Birefringent crystals (T and T') are used
to compensate for the temporal walk-off inside the down
conversion crystals, and birefringent crystals (c) compensate
spatial walk-off in modes $1$ and $3$. \textbf{b,} A two-photon
interferometer is used to project on $|\phi^-\rangle_{AC}$ for entanglement unlocking
between parties $B$ and $D$. The delay $\Delta\tau$ is adjusted for
optimum two-photon interference~\cite{Hong1987a}.
\label{fig:setup}}
\end{center}
\end{figure}

In any attempt to generate perfect Smolin states, the PPT property
will be very sensitive to imperfections in the state preparation and
low counting statistics in the experimental data. The main
reason is that the partial transpose of the density matrix of a perfect Smolin state is not full-rank. Introducing a source of
white noise leads to a full-rank matrix, whose PPT property is
more robust and thus better suited for experimental
demonstration. By changing the amount of white noise one generates
the family of noisy Smolin states
\begin{equation}
\label{eq:noisy_Smolin}
\rho_S(p)=(1-p)\rho_S+ p\frac{\mathcal{I}}{16},
\end{equation}
where $0\leq p\leq 1$ parameterizes the strength of the noise
applied, and $\mathcal{I}$ is the identity matrix. These states
are bound entangled for $0\leq p<2/3$, and fully separable for
$2/3\leq p\leq 1$~\cite{augusiak2006bound}. Entanglement can be ascertained by the use of an
entanglement witness, that is an observable $\mathcal{W}$ such
that $\Tr(\mathcal{W}\tau)<0$ for some entangled state $\tau$, while $\Tr(\mathcal{W}\rho)\geq0$ for all
separable states $\rho$. A suitable witness for these states is
$\cW=\mathcal{I}-\sum_{i=1}^3\sigma_i^{\otimes 4}$ like
in~\cite{amselem2009experimental}, where $\sigma_1,\sigma_2$
and $\sigma_3$ are the three Pauli matrices X, Y, Z. A short,
fully analytic derivation of this witness can be found in Appendix A.

\begin{figure*}
\begin{center}
\includegraphics[width=2 \columnwidth]{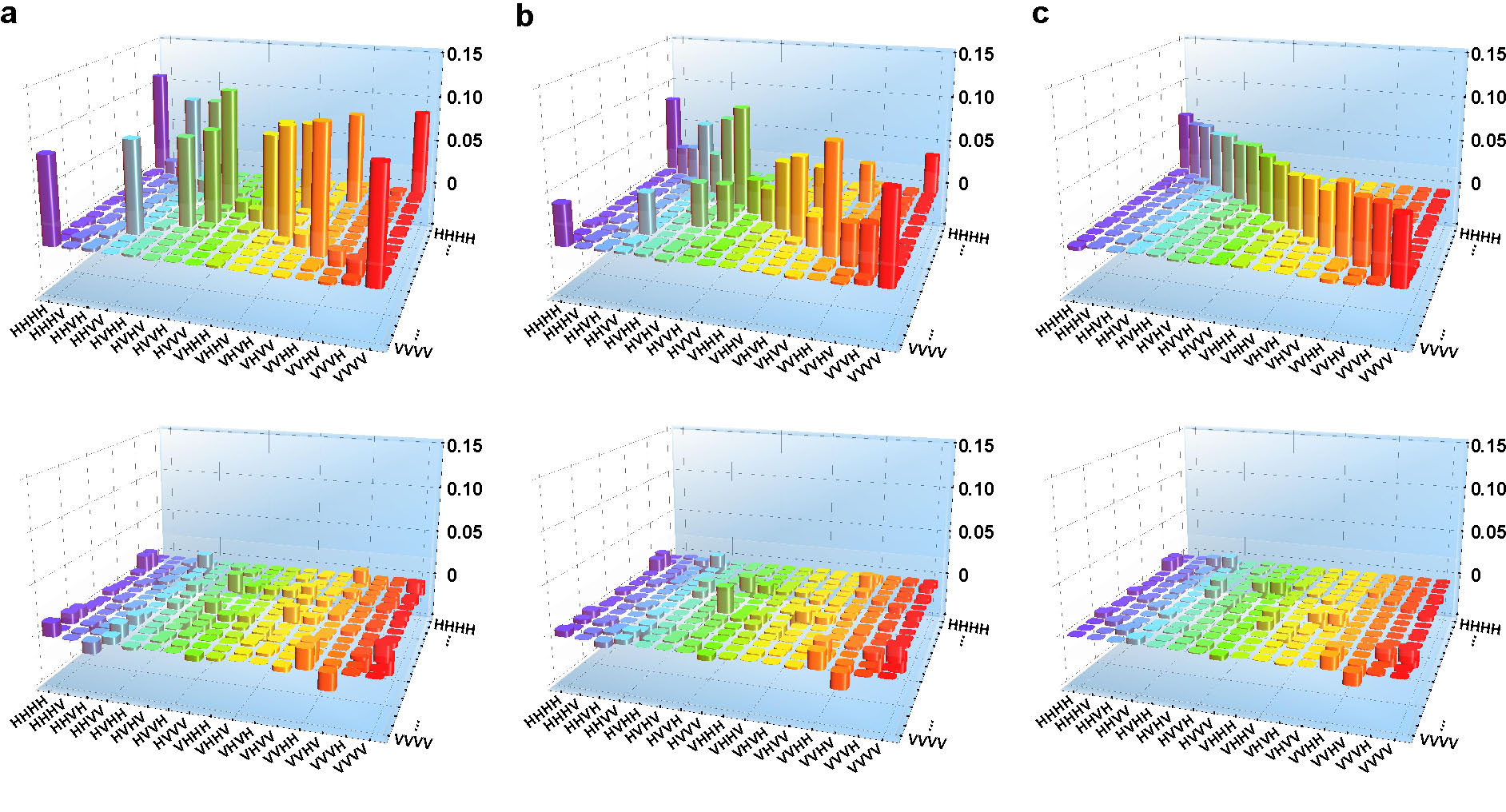}
\caption{\textbf{Experimentally measured density matrices of noisy Smolin states.} The upper row represents the
real part of the measured density matrices for noise levels,
$p$, of \textbf{(a)} $0.00$, \textbf{(b)} $0.49$ and \textbf{(c)} $1.00$ while the lower row
contains the corresponding imaginary parts. The fidelity with
the target state is \textbf{(a)} ($81.52\pm0.12$)\%, \textbf{(b)}
($96.83\pm0.05$)\% and \textbf{(c)} ($97.67\pm0.04$)\%, and the measured
witness and smallest PT eigenvalue are, respectively, \textbf{(a)}
$-1.269\pm0.006$ and $-0.0273\pm0.0006$, \textbf{(b)} $-0.159\pm0.008$
and $0.0069\pm0.0008$ and \textbf{(c)} $0.985\pm0.008$ and
$0.0301\pm0.0008$. From these values we conclude that \textbf{(a)} is
entangled but not PPT, \textbf{(c)} is PPT but not entangled, but \textbf{(b)}
is both entangled \textit{and} PPT, i.e. bound entangled.
\label{fig:dms}}
\end{center}
\end{figure*}

\begin{figure}
\begin{center}
\includegraphics[width=1 \columnwidth]{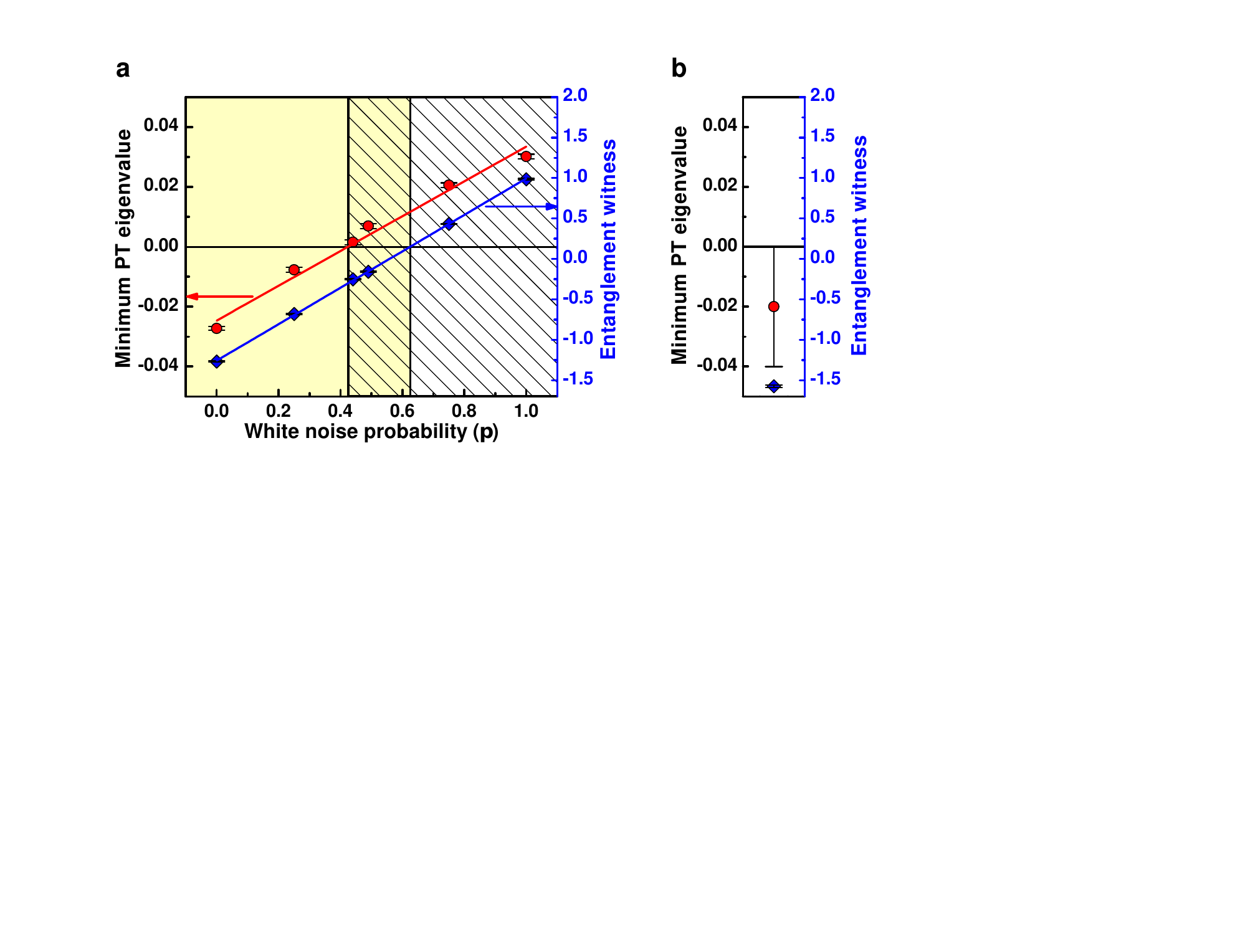}
\caption{\textbf{Experimental tests for bound entanglement.} Panel \textbf{(a)} shows the measured expectation value for the entanglement witness (blue
diamonds, right axis) and the minimum PT eigenvalue (red
circles, left axis) for various levels of white noise. The lines are best fits to the data and the error bars correspond to one standard deviation as determined by Monte-Carlo simulations based on Poisson distributions for the counts.
Each eigenvalue shown is the smallest
among all those calculated from the set of bipartite cuts
$(AB):(CD)$, $(AC):(BD)$, and $(AD):(BC)$. Using our experimental data, we find that our family of generated Smolin states is entangled within the shaded region and PPT in
the hatched region. In the overlapping region, the states are both
entangled and PPT, i.e. bound entangled. In particular, we successfully produce
a bound entangled state when the noise level is $p=0.49$; the
entanglement witness is $-0.159\pm 0.008$ and the minimum PT
eigenvalue is $0.0069\pm 0.0008$. For comparison,
\textbf{(b)} shows the results from
Ref.~\cite{amselem2009experimental}, where no additional noise was applied. Our state without additional noise is entangled but definitely not PPT, with a negative minimum PT eigenvalue similar to Ref.~\cite{amselem2009experimental}.
Our experimental results show
that a substantial amount of noise is required to turn this
value firmly positive.
\label{fig:results}}
\end{center}
\end{figure}

Smolin states can be prepared in the following way. Two sources
of entangled pairs produce a state of the form
$\ket{\phi^+}_{AB}\otimes\ket{\phi^+}_{CD}$, where
$|\phi^\pm\rangle = \frac{1}{\sqrt{2}}\left(|00\rangle
\pm|11\rangle\right)$. One then applies randomly, but with equal
weight, one of the rotations
$\sigma^\mu$, $\mu=0,\ldots,3$, with $\sigma^0$ the identity,
simultaneously to both entangled pairs, i.e., $\sigma^\mu_A|\phi^+\rangle_{AB}\otimes
\sigma^\mu_C|\phi^+\rangle_{CD}$. Different levels of white noise
can be created by choosing a probability, $p$, where the
rotations are applied in an uncorrelated fashion.

In our experiment, we take this approach to generate a
four-photon Smolin state of the form in equation~(\ref{eq:noisy_Smolin}). We use spontaneous parametric down
conversion to generate two photon pairs in four distinct
spatial modes, as shown in Fig.~\ref{fig:setup}a. Both pairs
are originally in the state $|\phi^+\rangle$, encoded in the
polarization of the photons, where $|0\rangle=|\SH\rangle$
and $|1\rangle=|\SV\rangle$ correspond to horizontal and
vertical polarization, respectively. We implement the unitaries, $\sigma^\mu$,
via a pair of liquid-crystal variable phase retarders (LCRs) in
each of the two sources. For more details see the Methods section.

To characterize our experimentally generated four-qubit states,
we perform quantum-state tomography on the polarized photons
with an over-complete set of measurements. We use an iterative
maximum-likelihood algorithm~\cite{Jezek2003a} to reconstruct
the density matrix. In the Methods section and the Appendix B, we
provide more details on our measurement and state reconstruction procedure.
In Fig.~\ref{fig:dms}, we show the real and imaginary part of the
reconstructed density matrices for three different levels of
white noise. For $p=0.00$, $p=0.49$ and $p=1.00$, the fidelity~\cite{Jozsa1994a} with the target state $\rho_S (p)$ is ($81.52\pm0.12$)\%,
($96.83\pm0.05$)\% and ($97.67\pm0.04$)\%, respectively. The
error bars are estimated from Monte-Carlo simulations with
$500$ iterations each. Fidelities for the additional noise
levels $p=0.25$, $p=0.44$ and $p=0.75$ are given in Table~\ref{table:WitnessCounts}.

\begin{figure}
\begin{center}
\includegraphics[width=1 \columnwidth]{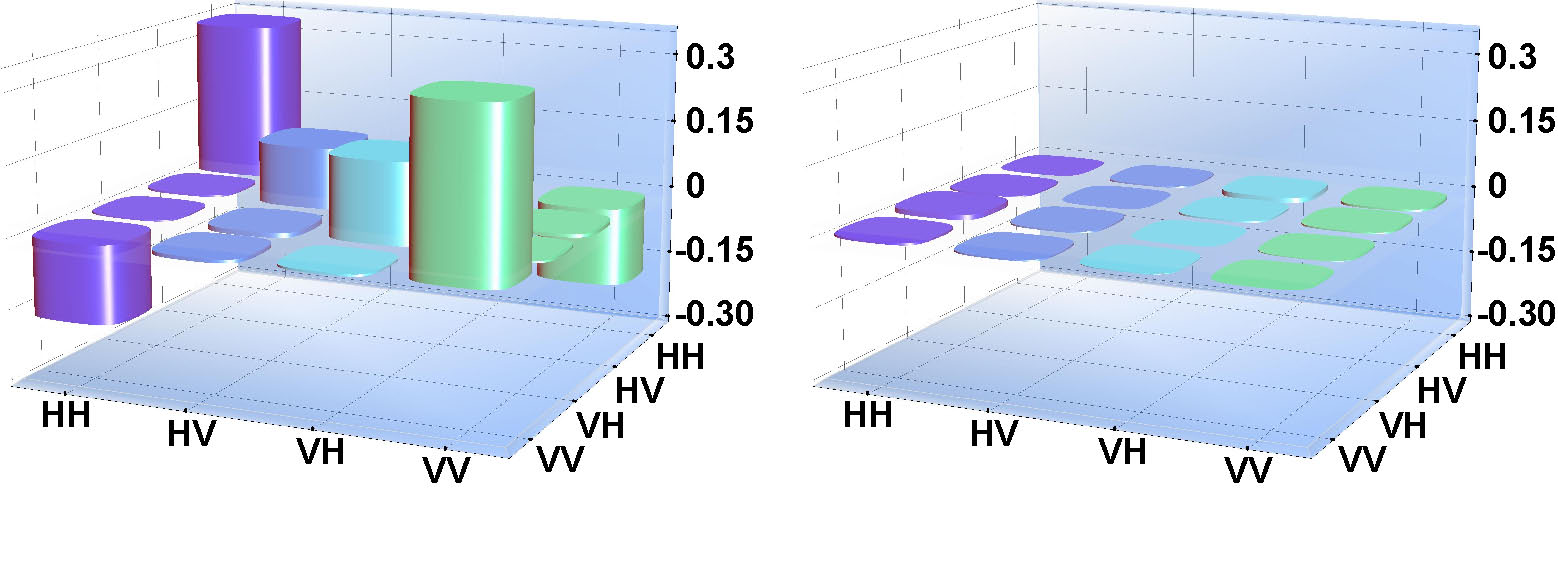}
\caption{\textbf{Unlocking of entanglement from a bound entangled state.} By performing a joint measurement on the
qubits of parties $A$ and $C$, specifically by projecting them
on the state $\ket{\phi^-}_{AC}$ using the Bell measurement
shown in Fig.~\ref{fig:setup}b, entanglement is unlocked
between parties $B$ and $D$. The density matrix of the
resulting state shared by $B$ and $D$ is reconstructed via an
over-complete set of tomography measurements. Here, we
use a noise level of $p=0.49$ so that the four-qubit state
prepared is bound entangled. The reconstructed density matrix of
the unlocked entangled state (real part on the left, imaginary
part on the right) is clearly entangled with a negative minimum
PT eigenvalue of $-0.0160\pm0.0035$ and a tangle of
$0.00105\pm0.00046$. We achieve a fidelity of
($99.45\pm0.05$)\% with the state expected, given the
reconstructed four-qubit density matrix of Fig.~\ref{fig:dms}b.
\label{gif:activation}}
\end{center}
\end{figure}

In order to test whether our prepared state is bound entangled,
we directly measure the entanglement witness and determine if the
reconstructed density matrix is PPT. When no white noise is
added to our Smolin state, only one of these conditions is
fulfilled: the state is entangled but the minimum partial-transpose (PT) eigenvalue
is negative (see Fig.~\ref{fig:results}a). The latter is
defined as the minimum of all eigenvalues of the
partially transposed density matrix with respect to all two-two bipartite
cuts. The magnitude of the negative eigenvalue is similar to
that presented in~\cite{amselem2009experimental}, which we
provide in Fig.~\ref{fig:results}b for comparison. To
reach PPT one can add white noise. For each level of white noise, Fig.~\ref{fig:results}a
shows the directly measured entanglement witness and the minimum PT
eigenvalue of our reconstructed state. As the noise
probability increases, so do both the witness expectation value
and the minimum PT eigenvalue. Our state is bound entangled in
the region where the witness is negative while the minimum PT
eigenvalue is non-negative. In particular, for $p=0.49$ the
measured value for the witness is $-0.159\pm0.008$ (see Table~\ref{table:SmolinFidelities}) and the minimum PT eigenvalue is
$0.0069\pm0.0008$. These values satisfy \emph{both} conditions
required for bound entanglement by a wide margin.

Although the entanglement of a Smolin state (equation~(\ref{eqn:smolin})) is undistillable
using only LOCC, it can be unlocked between any two of the
parties when joint operations between the other two parties are
allowed~\cite{smolin2001four}. Indeed, by performing a Bell
state measurement, two parties, e.g. $A$ and $B$, can
find out which Bell pair they share. They can communicate that
information to $C$ and $D$, who will share the same Bell state.
In this case, $C$ and $D$ will end up sharing a maximally
entangled state although, as discussed, no entanglement could
have been distilled between them via four-party LOCC. In the case of the family of Smolin states described in equation~(\ref{eq:noisy_Smolin}) a Bell measurement on any two parties will lead to the preparation of a Werner state~\cite{Werner89} in the other two. The resulting Werner state will be entangled for $p<\frac{2}{3}$, i.e. as long as the Smolin state is entangled.

To demonstrate entanglement unlocking, we keep the
noise level at $p=0.49$, immediately after obtaining the results above, and we feed the photons of parties $A$
and $C$, i.e., one from each source, into the Bell measurement
depicted in Fig.~\ref{fig:setup}b. This performs a projection measurement of modes $A$ and $C$ onto the Bell state $|\phi^-\rangle$. An over-complete set of two-qubit tomography measurements is
performed on the qubits of parties $B$ and $D$, yielding the
counts reported in Table~\ref{table:ActivationCounts}. The
reconstructed density matrix (see Fig.~\ref{gif:activation}) has
a fidelity of $(99.45\pm 0.05)\%$ with the one expected from
the experimental bound entangled state, assuming a perfect Bell
measurement. Its tangle is positive, $0.00105\pm0.00046$, and the minimum
 eigenvalue of the partially transposed state is negative, $-0.0160\pm0.0035$;
 %A positive tangle and non-positive partially transposed state both
 this confirms that we have successfully
unlocked entanglement between parties $B$ and $D$.

We demonstrated experimental bound entanglement, for the
first time convincingly satisfying its two defining criteria:
entanglement and undistillability. To achieve the latter
property, we added sufficient white noise to clearly fulfill the PPT
criterion while still maintaining non-separability.
Without additional noise, our Smolin state is non-PPT by an amount that is relatively small, but statistically significant, and we must add a \textit{substantial} amount of noise---almost $50\%$---to turn it PPT. Clearly, it is difficult to achieve the critical PPT condition without completely losing entanglement in experimentally produced Smolin states.
Once we achieved the preparation of bound entanglement, we demonstrated entanglement unlocking, realizing all of the conceptually important characteristics of a Smolin state. Our results open the door to applications of bound entanglement in experimental quantum information science.

\begin{acknowledgements}
We acknowledge financial support from NSERC, CFI, Ontario MRI ERA and OCE. We thank C. Ryan, O. G\"{u}hne,
R. Laflamme and R. Prevedel for valuable discussions.
\end{acknowledgements}

\section*{Methods}

Two polarized photon pairs are generated using $\beta$-BaB$_2$O$_4$ (BBO) nonlinear crystals phase-matched for type-I parametric down-conversion. Each photon source consists of two orthogonally oriented, $1$mm thick crystals~\cite{Kwiat1999a, Lavoie2009a}. The pump laser beam is a $830$mW, (FWHM$=1.6$~nm, centered at $395$nm) generated in a $2$mm thick Bismuth-Borate (BiBO) crystal by second harmonic generation, from a $2.85$W Ti:sapphire pulsed laser (FWHM$=9.8$nm, centered at $790.4$nm). Birefringent crystals ($1$mm thick BiB$_3$O$_6$ cut at $\theta = 152.6^\circ$ and $\phi=0^\circ$) are added in modes $1$ and $3$ shown in Fig.~\ref{fig:setup}a to compensate for the transverse walk-off occurring inside the down-conversion crystals. The longitudinal walk-off is compensated using a $0.5$mm and $2$mm quartz plus a $1$mm $\alpha$-BBO crystal before the first SPDC source, and a $2$mm $\alpha$-BBO and a $1$mm quartz crystal between the two sources. The photons are coupled into single mode fibers after going through an interference filter with a $3$nm bandwidth centered at $790$nm. We initially adjust the phase with a quarter-wave plate (QWP) in modes $2$ and $4$ such that each source produces $\ket{\phi^+}$. The average singles rate produced is $250$kHz with an average coincidence rate of $36$kHz, when each analyzer is set to $|\SH\rangle$ and monitoring both outputs of the polarising beam-splitter (PBS).

We use one pair of LCRs per source as depicted in Fig.~\ref{fig:setup}a. Each pair is composed of one LCR that can be ``off'', applying the identity, or ``on'', acting as an X and one LCR acting as either the identity or Z when turned off or on, respectively. When they are both on, they perform an XZ. For each of the sources and for each combination of LCR states, Table~\ref{table:sourcecharacterisation} lists the fidelity and the tangle of the state resulting from the LCRs acting on the initial $|\phi^+\rangle$ state. At any given time, the state of the LCRs is set by a computer using a pseudo-random number generator operating at a rate of $10$~Hz.

The polarization of each qubit is analyzed via a half-wave plate, and a QWP followed by a PBS (see Fig.~\ref{fig:setup}), where we monitor both output modes of the PBS. For quantum state tomography, we perform an over-complete set of standard polarization measurements $|\SH\rangle\slash|\SV\rangle=|0\rangle\slash|1\rangle$, $|\SD\rangle\slash|\SA\rangle=\frac{1}{\sqrt{2}}(|\SH\rangle\pm|\SV\rangle)$, $\ket{\SR}=\frac{1}{\sqrt{2}}(|\SH\rangle + i|\SV\rangle)$, and  $\ket{\SL}=\frac{1}{\sqrt{2}}(|\SH\rangle - i|\SV\rangle)$. The counts for each measurement setting are four-fold coincidence detection events. We reconstruct our density matrices using an iterative maximum-likelihood algorithm~\cite{Jezek2003a}. A detailed description of our measurement technique is given in the Appendix B.

In our entanglement-unlocking experiment, we use single-mode fibers to feed the photons of parties $A$ and $C$ into a two-photon interferometer as depicted in Fig.~\ref{fig:setup}b. Polarization controllers assure that polarizations $\ket{\SH}$ and $\ket{\SV}$ are maintained within the fibers. After passing through the first PBS in the interferometer, both photons are projected on the state $\ket{\SD}$ using HWPs and PBSs. If the time delay $\Delta\tau$ is set such that two-photon interference occurs~\cite{Hong1987a}, a coincidence detection event between detectors D1 and D2 can only occur if the photons are in a state of the form $\frac{1}{\sqrt{2}}\left(\ket{\SH \SH}+\mathrm{e}^{i\chi}\ket{\SV \SV}\right)$. By using a QWP, we set the phase, $\chi$, such that a coincidence detection event signals a projection of the two-photon state on $\ket{\phi^-}$.

\appendix

\section{An entanglement witness for the Smolin state}
We consider an entanglement witness $\mathcal{W}$ such that a negative expectation value $\langle\mathcal{W}\rangle=\Tr(\rho \mathcal{W})<0$ is sufficient to exclude that the prepared state $\rho$ is in the set $\mathcal{S}$ of mixed states that are a convex combination of pure states with one party disentangled from the others, e.g. $\ket{\alpha}_{A}\otimes\ket{\phi}_{BCD}$. In such a convex combination, the disentangled party may differ from pure state to pure state. Of course, $\mathcal{S}$ is a superset of the set of completely separable states.

The witness reads $\mathcal{W}=\mathcal{I}-\sum_{i=1}^3\sigma_i^{\otimes 4}$.
There are a number of ways to single out $\mathcal{W}$ as an appropriate witness; e.g., in \cite{amselem2009experimental} the derivation was based on the stabilizer formalism and numerical optimization. Here we present a derivation that is completely analytical and uses the geometric approach of \cite{pittinger_rubin2002}.

It will be sufficient to prove that the state $\rho_S(2/3)$, which is fully separable \cite{augusiak2006bound}, is the closest to $\rho_S$ in the set $\mathcal{S}$. Here, closeness is defined with respect to the Hilbert-Schmidt norm $\|X\|_{\rm HS}=\sqrt{\Tr(X^\dagger X)}$. To this aim, using Proposition 5.1 in~\cite{pittinger_rubin2002}, one derives that it is sufficient to check that $\max_{\tau\in\mathcal{S}}\Tr(\rho_S\tau)\leq 1/8$.  We have
%\begin{widetext}
\begin{multline}
\max_{\tau\in\mathcal{S}}\Tr(\rho_S\tau)\\
\begin{aligned}
&=\max_{|\alpha\rangle_A|\phi\rangle_{BCD}}\Tr(\rho_S \proj{\alpha}_A\otimes \proj{\phi}_{BCD})\\
&=\max_{|\alpha\rangle_A|\phi\rangle_{BCD}}\Tr(\rho^{\Gamma_A}_S (\proj{\alpha})^T_A\otimes \proj{\phi}_{BCD})\\
&=\frac{1}{8}\left[1-4\min_{|\alpha\rangle_A|\phi\rangle_{BCD}}\Tr(\rho_S \proj{\alpha}_A\otimes \proj{\phi}_{BCD})\right]\\
&=\frac{1}{8},
\end{aligned}
\end{multline}
%\end{widetext}
where $\rho_S^{\Gamma_A}$ denotes the partial transpose of $\rho_S$ with respect to $A$. The first equality comes from the symmetry of $\rho_S$ and the convexity of $\mathcal{S}$, so that it is sufficient to consider pure states $|\alpha\rangle_A|\phi\rangle_{BCD}$ in the maximization; the second equality from the identity $\Tr(XY)=\Tr(X^{\Gamma_A} Y^{\Gamma_A})$; the third equality from $\rho^{\Gamma_A}= \frac{1}{8}(\sigma_2\otimes \sigma_0^{\otimes 3}) \frac{1}{8}\left(\mathcal{I}-4\rho_S\right)  (\sigma_2\otimes \sigma_0^{\otimes 3})$ and the fact that $\sigma_2(\proj{\alpha})^T_A\sigma_2$ is also a pure state. Finally, the minimum in the third line is easily seen to vanish.

Having established that $\rho_S(2/3)$ is the nearest state to $\rho_S$ in $\mathcal{S}$, according to Theorem 6.1 in~\cite{pittinger_rubin2002} one can construct a witness for $\rho_S$ as $\tilde{\mathcal{W}}=c_0\mathcal{I}+\rho_S(2/3)-\rho_S$, with $c_0=\Tr(\rho_S(2/3)(\rho_S-\rho_S(2/3))=1/24$, so that $\tilde{\mathcal{W}}\propto \mathcal{I}-\sum_{i=1}^3\sigma_i^{\otimes 4}=\mathcal{W}$. As $\rho_S(2/3)$ has full rank, Theorem 6.1 in~\cite{pittinger_rubin2002} also assures that $\mathcal{W}$ is optimal, i.e., there is no witness $\mathcal{W'}$ that detects all the states detected by $\mathcal{W}$ and some more.

\section{Quantum state tomography of Smolin states}
We perform an over-complete set of tomographic measurements for each level of white noise in order to reconstruct the density matrices of the states prepared. The measurement for each of the four qubits is chosen from a set of six states: $\ket{\SH}$, $\ket{\SV}$, $\ket{\SD}$, $\ket{\SA}$, $\ket{\SR}$, and $\ket{\SL}$, resulting in an overall set of $6^4 = 1296$ measurements. Each qubit measurement is implemented via a half-wave plate (HWP), followed by a quarter-wave plate (QWP) and a polarising beam splitter (PBS). By monitoring both PBS outputs simultaneously one can reduce the number of measurements to $3^4 = 81$ but this approach requires the two outputs of each PBS to have equal coupling efficiencies for all analyzer waveplate settings. Even if one tried to balance the coupling efficiencies, imperfections would lead to systematic errors in the measured expectation values, and balancing the coupling efficiencies is very susceptible to long-time drifts in the setup. We choose a more accurate and less sensitive approach as described in the following. Let us assume that we want to measure a qubit in the $\SH/\SV$ basis. First, we choose the angles of the HWP and QWP such that the path transmitted through the PBS corresponds to $\ket{\SH}$, the reflected one to $\ket{\SV}$. We record these counts, and then we repeat the measurement but with the waveplates oriented such that the transmitted path is $\ket{\SV}$, and the reflected one is $\ket{\SH}$. By adding the respective counts for these two cases, we automatically average over any imbalance in the coupling efficiencies of the two outputs of the PBS. The approach is the same for measuring in any other basis, and for each basis we have to average over two settings. For four qubits we have to average over $16$ settings, resulting in a total of $1296$ settings, but for each of them we get approximately $16$ times the number of counts we would expect when monitoring only the transmitted paths of our PBSs. In other words, using the counts from all PBS outputs this way effectively increases the measurement time per setting by a factor of $2^N$, where $N$ is the number of two-output qubit analyzers involved.

Because of the high number of measurement settings in each tomographic scan, we take additional precautions in order to prevent long-time drifts from influencing our measurement results. For each setting we record data for $5$s, and we repeat the full set of measurements $10$ times. The order in which the settings are measured is random for every loop we perform. We add together the counts of all loops such that the resulting counts correspond to a measurement time of $10\times 5 = 50$s per setting. Taking into account the additional factor of $16$ due to our use of all PBS outputs, the effective overall measurement time is $800$s per setting.

The method we use for the reconstruction of the density matrix is an iterative maximum likelihood technique introduced by Je\v{z}ek et al.~\cite{Jezek2003a}. We compared its results with those of another popular technique that poses the search for the maximum-likelihood density matrix as a semidefinite program \cite{Doherty2009a}, which we solve using CSDP (a C Library for Semi-Definite Programming). The latter method yields a global optimum for the density matrix but is in general significantly slower than the iterative maximum likelihood technique. For comparison we performed tomography of the counts measured for a noise level of $49$\% using both techniques. The fidelity between the reconstructed density matrices using the semidefinite maximum-likelihood technique and the iterative maximum-likelihood algorithm is $1.0000$.  In case of the iterative technique we performed $500$ iterations, in case of the semidefinite technique we performed $90$.  Using the semidefinite technique we calculate the minimum eigenvalue of the partial transpose of the reconstructed density matrix with respect to all two-two partitions (short: minimum PT eigenvalue) to be $0.007\pm0.001$, compared to the $0.0069\pm 0.0008$ we get using the iterative algorithm. These values are in good agreement.

\section{Minimum eigenvalues and the entanglement witness}

Before each loop in the tomographic measurements we performed a direct measurement of the entanglement witness. Again, we sum the counts of all $10$ loops and use these results for the witness at each white-noise level. To measure the witness we perform $48$ measurements. These correspond to the $3\times 16$ measurements for measuring the expectation values of $\sigma_1^{\otimes 4}$, for $\sigma_2^{\otimes 4}$, and for $\sigma_3^{\otimes 4}$. The expectation value of the witness is then given by $\langle\mathcal{W}\rangle = 1 - \sum^3_{i=1} \left\langle \sigma_i^{\otimes 4} \right\rangle$.

\begin{figure*}
  \begin{center}
    \includegraphics[width=2 \columnwidth]{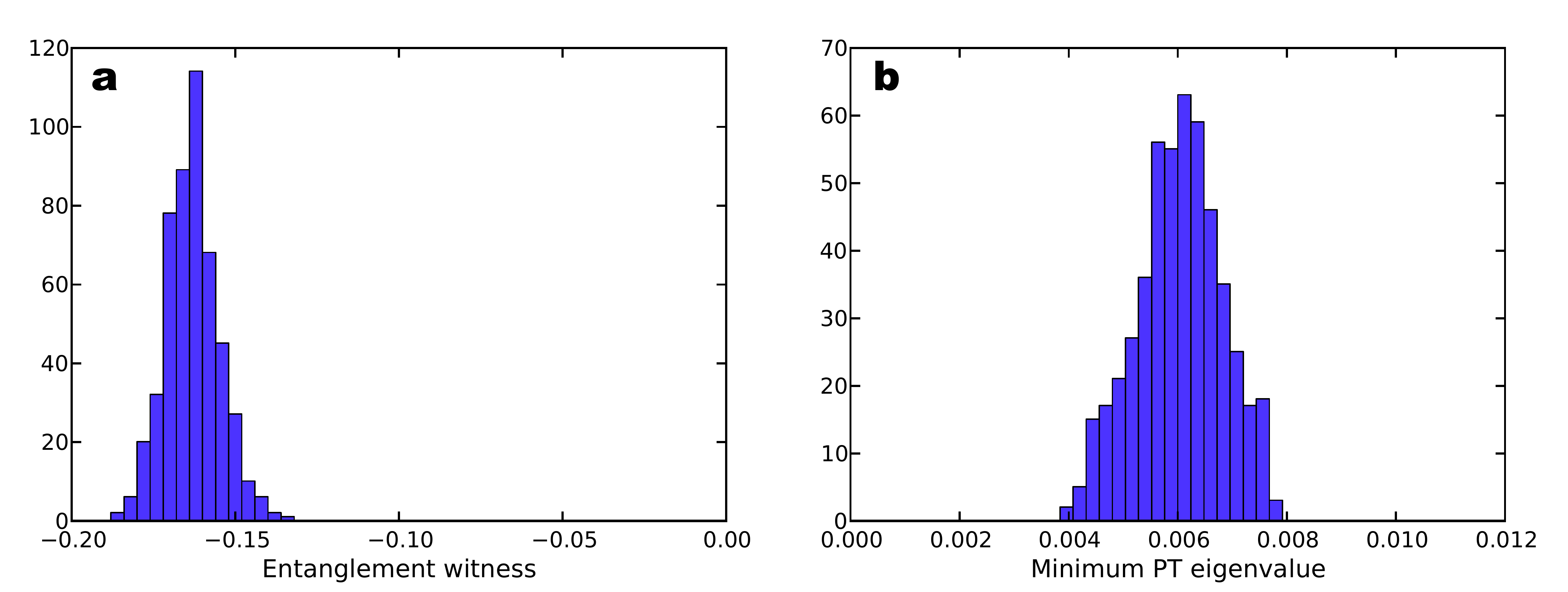}
    \caption{\textbf{Histograms for the entanglement witness and the minimum PT eigenvalue.} We used 500 Monte-Carlo iterations on the tomography data for $p=0.49$ to generate histograms for, \textbf{a}, the expectation value of the entanglement witness $\langle\mathcal{W}\rangle$ and, \textbf{b}, the minimum PT eigenvalue. These histograms show that the expectation value of the entanglement witness is strictly negative, and that the minimum PT eigenvalue is strictly positive. }\label{fig:fig5}
  \end{center}
\end{figure*}

\begin{figure*}
  \begin{center}
    \includegraphics[width=2 \columnwidth]{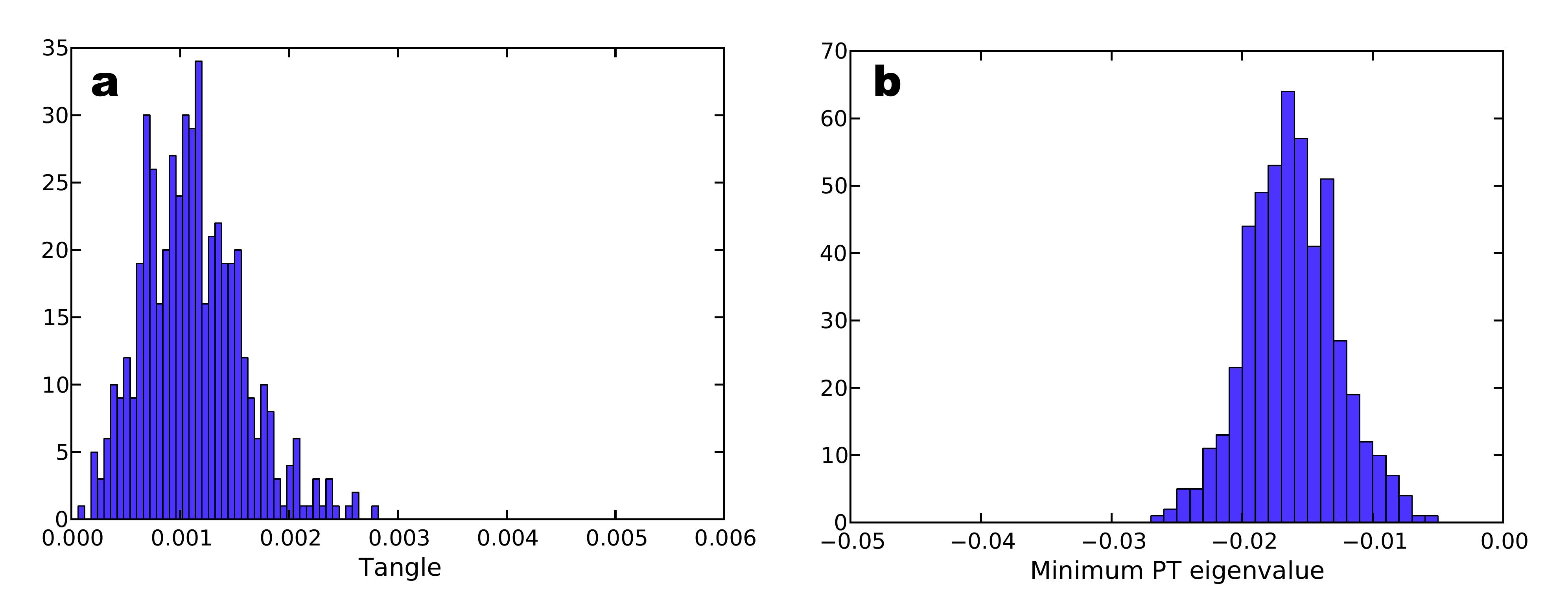}
    \caption{\textbf{Histograms for the tangle and minimum PT eigenvalue of unlocked entanglement.} Using the two-qubit tomography data for the qubits of parties $B$ and $D$ after projecting the two other qubits on $\ket{\phi^-}$, we perform 500 Monte-Carlo iterations. Panels \textbf{a} and \textbf{b} show histograms for the tangle and the minimum eigenvalue of the partially-transposed reconstructed density matrix, respectively. Here, we use minimum PT eigenvalue in the same meaning as we do for the four-qubit Smolin state although, of course, for two qubits there is only one partition with respect to which the partial transpose can be performed. Both histograms confirm that entanglement has been unlocked.}\label{fig:fig6}
  \end{center}
\end{figure*}

We give the counts for the witness measurement for a noise level of $49$\% as well as the expectation values for the three products of Pauli operators in Table~\ref{table:WitnessCounts}. Using these results we get an expectation value of $\langle\mathcal{W}\rangle = -0.159 \pm 0.008$. Table~\ref{table:SmolinFidelities} shows the witness values measured for various noise levels, and, for comparison, we also provide the witness values calculated from the reconstructed density matrices.

In Fig.~\ref{fig:fig5} we show histograms of all witness values and minimum PT eigenvalues for the 500 density matrices calculated in the course of the Monte-Carlo simulation performed on the counts recorded for a noise level of $49$\%. These histograms further illustrate that the state fulfills the criteria for bound entanglement.

For our entanglement-unlocking results, we provide the count rates measured for the tomography of the two-qubit density matrix shown in Fig.~\ref{gif:activation} of the main text in Table~\ref{table:ActivationCounts}. The histograms in Fig.~\ref{fig:fig6} show that the tangle is greater than zero, and that the minimum eigenvalue of the partially transposed density matrix is negative. The histograms comprise the values calculated in $500$ iterations of a Monte-Carlo simulation using the measured data. They provide evidence for entanglement between the remaining photons, i.e., we have successfully unlocked entanglement from an initially bound entangled state.

\section{Characterization of the down-conversion sources}
Depending on the state of the LCRs, different unitaries are applied to the states generated by our sources. Both sources are aligned to produce $\ket{\phi^+} = \frac{1}{\sqrt{2}}(\ket{\SH\SH} + \ket{\SV\SV})$ states. One of the two LCRs in each source performs the identity ($\sigma_0$) when in the off state, and it performs a Pauli X when in the on state. The second LCR performs $\sigma_0$ when off and a Pauli Z when on. The combination of the two LCRs determines the overall unitary applied to the entangled state.

In order to test the performance of the two sources, we perform two-qubit tomographic scans  on the photon pairs emitted by the two sources. We do these tomography measurements for all four states of the LCRs (off-off, off-on, on-off, on-on) for each of the sources. For each set of these tomographic measurements we perform a Monte-Carlo simulation with $500$ iterations, using the measured counts as the means of Poissonian distributions; these two-qubit measurements were taken immediately before the tomography of the Smolin state with $49$\% noise, and are given in Table~\ref{table:sourcecharacterisation}. The values for fidelity and tangle are typical for the performance of our down-conversion sources.

\begin{table*}
 {\small\begin{tabular}{lcclcclcc}
    \hline\hline
    states & counts & \hspace{0.5em} & states & counts & \hspace{0.5em} & states & counts & \hspace{0.5em}\\[0.3ex]\hline
$\ket{\SH,\SH,\SH,\SH}$ & 3834 & & $\ket{\SD,\SD,\SD,\SD}$ & 3687 & & $\ket{\SR,\SR,\SR,\SR}$ & 3810\\[0.3ex]
$\ket{\SH,\SH,\SH,\SV}$ & 1983 & & $\ket{\SD,\SD,\SD,\SA}$ & 1751 & & $\ket{\SR,\SR,\SR,\SL}$ & 1833\\[0.3ex]
$\ket{\SH,\SH,\SV,\SH}$ & 1760 & & $\ket{\SD,\SD,\SA,\SD}$ & 1801 & & $\ket{\SR,\SR,\SL,\SR}$ & 1813\\[0.3ex]
$\ket{\SH,\SH,\SV,\SV}$ & 4344 & & $\ket{\SD,\SD,\SA,\SA}$ & 3863 & & $\ket{\SR,\SR,\SL,\SL}$ & 3825\\[0.3ex]
$\ket{\SH,\SV,\SH,\SH}$ & 1531 & & $\ket{\SD,\SA,\SD,\SD}$ & 1658 & & $\ket{\SR,\SL,\SR,\SR}$ & 1801\\[0.3ex]
$\ket{\SH,\SV,\SH,\SV}$ & 4050 & & $\ket{\SD,\SA,\SD,\SA}$ & 4125 & & $\ket{\SR,\SL,\SR,\SL}$ & 4021\\[0.3ex]
$\ket{\SH,\SV,\SV,\SH}$ & 3627 & & $\ket{\SD,\SA,\SA,\SD}$ & 4276 & & $\ket{\SR,\SL,\SL,\SR}$ & 4029\\[0.3ex]
$\ket{\SH,\SV,\SV,\SV}$ & 1641 & & $\ket{\SD,\SA,\SA,\SA}$ & 1738 & & $\ket{\SR,\SL,\SL,\SL}$ & 1742\\[0.3ex]
$\ket{\SV,\SH,\SH,\SH}$ & 1500 & & $\ket{\SA,\SD,\SD,\SD}$ & 1524 & & $\ket{\SL,\SR,\SR,\SR}$ & 1711\\[0.3ex]
$\ket{\SV,\SH,\SH,\SV}$ & 4174 & & $\ket{\SA,\SD,\SD,\SA}$ & 3891 & & $\ket{\SL,\SR,\SR,\SL}$ & 3745\\[0.3ex]
$\ket{\SV,\SH,\SV,\SH}$ & 3608 & & $\ket{\SA,\SD,\SA,\SD}$ & 4023 & & $\ket{\SL,\SR,\SL,\SR}$ & 3954\\[0.3ex]
$\ket{\SV,\SH,\SV,\SV}$ & 1617 & & $\ket{\SA,\SD,\SA,\SA}$ & 1608 & & $\ket{\SL,\SR,\SL,\SL}$ & 1842\\[0.3ex]
$\ket{\SV,\SV,\SH,\SH}$ & 3737 & & $\ket{\SA,\SA,\SD,\SD}$ & 4094 & & $\ket{\SL,\SL,\SR,\SR}$ & 3845\\[0.3ex]
$\ket{\SV,\SV,\SH,\SV}$ & 1909 & & $\ket{\SA,\SA,\SD,\SA}$ & 1754 & & $\ket{\SL,\SL,\SR,\SL}$ & 1905\\[0.3ex]
$\ket{\SV,\SV,\SV,\SH}$ & 1709 & & $\ket{\SA,\SA,\SA,\SD}$ & 1843 & & $\ket{\SL,\SL,\SL,\SR}$ & 1963\\[0.3ex]
$\ket{\SV,\SV,\SV,\SV}$ & 4217 & & $\ket{\SA,\SA,\SA,\SA}$ & 3990 & & $\ket{\SL,\SL,\SL,\SL}$ & 3970\\[0.3ex]\hline
exp. values & $\EV{\sigma_z^{\otimes 4}}=0.3966\pm 0.0075$ & & & $\EV{\sigma_x^{\otimes 4}}=0.4005\pm0.0043$ & & & $\EV{\sigma_y^{\otimes 4}}=0.3621\pm 0.0043$ \\[0.3ex]\hline\hline
 \end{tabular}}
 \caption{\textbf{Counts for the measurement of the entanglement witness.} The counts are four-fold coincidences measured to determine the expectation value of the entanglement witness, $\langle\mathcal{W}\rangle = 1 - \sum^3_{i=1} \left\langle \sigma_i^{\otimes 4} \right\rangle$, for a noise level of $p=0.49$. We integrated over $60$s ($6$s in each of the ten loops) per measurement setting. Because we use all PBS outputs, the effective measurement time is $16\times60 = 960$s per setting.\label{table:WitnessCounts}}
\end{table*}

\begin{table*}
 \begin{tabular}{cccccccc}
    \hline\hline
 & \hspace{0.5em} & \multicolumn{6}{c}{noise level (in \%)}\\[0.3ex]\cline{3-8}
 & & 0 & 25 & 44 & 49 & 75 & 100 \\[0.3ex]\hline
fidelity & & $81.52\pm 0.12$ & $95.11\pm 0.07$ & $96.09\pm 0.06$ & $96.83\pm 0.05$ & $97.40\pm 0.03$ & $97.67\pm 0.04$\\[0.3ex]
min. EV & & $ -0.0273\pm 0.0006$ & $-0.0077\pm 0.0008$ & $0.0015\pm0.0008$ & $0.0069\pm 0.0008$ & $0.0206\pm 0.0008$ & $0.0301\pm 0.0008 $\\[0.3ex]
$\EV{\mathcal{W}_{\mathrm{est}}}$ & & $-1.261\pm 0.005$ & $-0.673\pm 0.007$ & $-0.239\pm 0.008$ & $-0.164\pm 0.008$ & $0.419\pm 0.008$ & $1.017\pm 0.009$ \\[0.3ex]
$\EV{\mathcal{W}_{\mathrm{sum}}}$ & & $-1.269\pm 0.006$ & $-0.682\pm 0.007$ & $-0.253\pm 0.007$ & $-0.159\pm 0.008$ & $0.432\pm 0.007$ & $0.985\pm 0.008$  \\[0.3ex]\hline\hline
 \end{tabular}
 \caption{\textbf{Characterization of the experimentally prepared noisy Smolin states.} For various levels of white noise, we show the fidelity of the reconstructed density matrix with the theoretically expected density matrix of a noisy Smolin state. With ``min. EV'' we denote the minimum PT eigenvalue. At every noise level, we give two values for the expectation value of the entanglement witness. The value we estimate from the reconstructed density matrix is $\EV{\mathcal{W}_{\mathrm{est}}}$, whereas we denote the value that we directly measured experimentally (see Table~\ref{table:WitnessCounts}) as $\EV{\mathcal{W}_{\mathrm{sum}}}$. These values are in good agreement. In the latter case we calculate the error using error propagation and assuming Poissonian errors for the measured counts. The values for the other quantities (fidelity, minimum PT eigenvalue and $\EV{\mathcal{W}_{est}}$) are calculated using the reconstructed density matrix,  and the errors are estimated via Monte-Carlo simulations with $500$ iterations each. \label{table:SmolinFidelities}}
\end{table*}

\begin{table*}
 \begin{center}
 \begin{tabular}{ccccccccccccccc}
    \hline\hline
& & & & \multicolumn{10}{c}{Photon D} \\[0.3ex]\cline{5-15}
& \hspace{0.2em} & & \hspace{0.5em} & $\ket{\SH}$ & \hspace{0.5em} & $\ket{\SV}$ & \hspace{0.5em} & $\ket{\SD}$ & \hspace{0.5em} & $\ket{\SA}$ & \hspace{0.5em} & $\ket{\SR}$ & \hspace{0.5em} & $\ket{\SL}$\\[0.3ex]\hline
 \multirow{6}{*}{\rotatebox{90}{Photon B}} & & $\ket{\SH}$ & & 4547 & & 1736 & & 3195 & & 3053 & & 3142 & & 3026\\[0.3ex]
 & & $\ket{\SV}$ & & 2274 & & 4626 & & 3463 & & 3544 & & 3539 & & 3464\\[0.3ex]
 & & $\ket{\SD}$ & & 3244 & & 3039 & & 1935 & & 4188 & & 3208 & & 3117\\[0.3ex]
 & & $\ket{\SA}$ & & 3261 & & 3185 & & 4277 & & 2098 & & 3385 & & 3325\\[0.3ex]
 & & $\ket{\SR}$ & & 3560 & & 3207 & & 3282 & & 3183 & & 4310 & & 2283\\[0.3ex]
 & & $\ket{\SL}$ & & 3266 & & 3308 & & 3062 & & 3176 & & 2204 & & 4230\\[0.3ex]\hline\hline
\end{tabular}
 \caption{\textbf{Counts for the two-qubit tomography measurement for our demonstration of entanglement unlocking.} We project the qubits of parties $A$ and $C$ on $\ket{\phi^-}$ and perform an over-complete set of tomography measurements on the qubits of parties $B$ and $D$. Each row rorresponds to a fixed setting for $B$, and each column to a fixed setting of $D$. All counts are four-fold coincidence events. We repeated the set of $36$ measurements $20$ times, and for each measurement setting we integrated over $60$s. Taking into account the additional factor of $4$ due to our use of all PBS outcomes, this results in an overall measurement time of $4\times20\times60=4800$s per setting.\label{table:ActivationCounts}}
 \end{center}
\end{table*}

\begin{table*}
 \begin{tabular}{ccccccccccc}
    \hline\hline
 & \hspace{0.3em} & & \hspace{0.5em} &  & \hspace{0.5em} & \multicolumn{2}{c}{fidelity} & \hspace{0.5em} & \multicolumn{2}{c}{tangle} \\[0.3ex]\cline{7-8}\cline{10-11}
LCR 1 & & LCR 2 & & ideal state & & source 1 & source 2 & & source 1 & source 2\\[0.3ex]\hline
$\sigma_0$ & & $\sigma_0$ & & $\ket{\phi^+}$ & & $95.77\pm0.02$\% & $96.34\pm0.02$\% & & $0.8709\pm0.0008$ & $0.8685\pm0.0007$ \\[0.3ex]
$\sigma_0$ & & $\sigma_3$ & & $\ket{\phi^-}$ & & $96.38\pm0.02$\% & $96.30\pm0.02$\% & & $0.8690\pm0.0007$ & $0.8698\pm0.0007$ \\[0.3ex]
$\sigma_1$ & & $\sigma_0$ & & $\ket{\psi^+}$ & & $96.58\pm0.02$\% & $96.49\pm0.02$\% & & $0.8705\pm0.0007$ & $0.8688\pm0.0006$ \\[0.3ex]
$\sigma_1$ & & $\sigma_3$ & & $\ket{\psi^-}$ & & $96.12\pm0.02$\% & $96.51\pm0.02$\% & & $0.8668\pm0.0007$ & $0.8664\pm0.0006$ \\[0.3ex]\hline\hline
% &
 \end{tabular}
 \caption{\textbf{Source characterization.} For each of the two sources we give the tangle of the state produced and its fidelity with the ideal entangled state we want to prepare ($\ket{\phi^{\pm}} = \frac{1}{\sqrt{2}}(\ket{\SH\SH}\pm\ket{\SV\SV})$ and $\ket{\psi^{\pm}} = \frac{1}{\sqrt{2}}(\ket{\SH\SV}\pm\ket{\SV\SH})$ are the four Bell states). Each value given is calculated from the density matrix reconstructed using an iterative maximum-likelihood technique on the results of an over-complete set of tomography measurements. The corresponding errors are calculated using a Monte-Carlo simulation with $500$ iterations.
 \label{table:sourcecharacterisation}}
\end{table*}


\clearpage



\begin{thebibliography}{10}
\expandafter\ifx\csname url\endcsname\relax
  \def\url#1{\texttt{#1}}\fi
\expandafter\ifx\csname urlprefix\endcsname\relax\def\urlprefix{URL }\fi
\providecommand{\bibinfo}[2]{#2}
\providecommand{\eprint}[2][]{\url{#2}}

\bibitem{Schroedinger1935a}
\bibinfo{author}{Schr{\" o}dinger, E.}
\newblock \bibinfo{title}{{D}ie gegenw{\" a}rtige {S}ituation in der
  {Q}uantenmechanik}.
\newblock \emph{\bibinfo{journal}{{D}ie {N}aturwissenschaften}}
  \textbf{\bibinfo{volume}{23}}, \bibinfo{pages}{807--812; 823--828; 844--849}
  (\bibinfo{year}{1935}).

\bibitem{HHHH09}
\bibinfo{author}{Horodecki, R.}, \bibinfo{author}{Horodecki, P.},
  \bibinfo{author}{Horodecki, M.} \& \bibinfo{author}{Horodecki, K.}
\newblock \bibinfo{title}{Quantum entanglement}.
\newblock \emph{\bibinfo{journal}{Rev. Mod. Phys.}}
  \textbf{\bibinfo{volume}{81}}, \bibinfo{pages}{865--942}
  (\bibinfo{year}{2009}).

\bibitem{ekert1991quantum}
\bibinfo{author}{Ekert, A.~K.}
\newblock \bibinfo{title}{Quantum cryptography based on {B}ell's theorem}.
\newblock \emph{\bibinfo{journal}{Phys. Rev. Lett.}}
  \textbf{\bibinfo{volume}{67}}, \bibinfo{pages}{661--663}
  (\bibinfo{year}{1991}).

\bibitem{bennett1993teleporting}
\bibinfo{author}{Bennett, C.~H.} \emph{et~al.}
\newblock \bibinfo{title}{{T}eleporting an unknown quantum state via dual
  classical and {E}instein-{P}odolsky-{R}osen channels}.
\newblock \emph{\bibinfo{journal}{Phys. Rev. Lett.}}
  \textbf{\bibinfo{volume}{70}}, \bibinfo{pages}{1895--1899}
  (\bibinfo{year}{1993}).

\bibitem{Mattle1996a}
\bibinfo{author}{Mattle, K.}, \bibinfo{author}{Weinfurter, H.},
  \bibinfo{author}{Kwiat, P.~G.} \& \bibinfo{author}{Zeilinger, A.}
\newblock \bibinfo{title}{{D}ense {C}oding in {E}xperimental {Q}uantum
  {C}ommunication}.
\newblock \emph{\bibinfo{journal}{Phys. Rev. Lett.}}
  \textbf{\bibinfo{volume}{76}}, \bibinfo{pages}{4656--4659}
  (\bibinfo{year}{1996}).

\bibitem{Bouwmeester1997a}
\bibinfo{author}{Bouwmeester, D.} \emph{et~al.}
\newblock \bibinfo{title}{{E}xperimental quantum teleportation}.
\newblock \emph{\bibinfo{journal}{Nature}} \textbf{\bibinfo{volume}{390}},
  \bibinfo{pages}{575--579} (\bibinfo{year}{1997}).

\bibitem{Jennewein2000a}
\bibinfo{author}{Jennewein, T.}, \bibinfo{author}{Simon, C.},
  \bibinfo{author}{Weihs, G.}, \bibinfo{author}{Weinfurter, H.} \&
  \bibinfo{author}{Zeilinger, A.}
\newblock \bibinfo{title}{{Q}uantum {C}ryptography with {E}ntangled {P}hotons}.
\newblock \emph{\bibinfo{journal}{Phys. Rev. Lett.}}
  \textbf{\bibinfo{volume}{20}}, \bibinfo{pages}{4729--4732}
  (\bibinfo{year}{2000}).

\bibitem{Nielsen2000a}
\bibinfo{author}{Nielsen, M.} \& \bibinfo{author}{Chuang, I.}
\newblock \emph{\bibinfo{title}{Quantum Computation and Quantum Information
  Theory}} (\bibinfo{publisher}{Cambridge Univ. Press},
  \bibinfo{address}{Cambridge}, \bibinfo{year}{2000}).

\bibitem{Bennett1996b}
\bibinfo{author}{Bennett, C.~H.}, \bibinfo{author}{Bernstein, H.~J.},
  \bibinfo{author}{Popescu, S.} \& \bibinfo{author}{Schumacher, B.}
\newblock \bibinfo{title}{{C}oncentrating partial entanglement by local
  operations}.
\newblock \emph{\bibinfo{journal}{Phys. Rev. A}} \textbf{\bibinfo{volume}{53}},
  \bibinfo{pages}{2046--2052} (\bibinfo{year}{1996}).

\bibitem{bennett1996mixed}
\bibinfo{author}{Bennett, C.~H.}, \bibinfo{author}{DiVincenzo, D.~P.},
  \bibinfo{author}{Smolin, J.~A.} \& \bibinfo{author}{Wootters, W.~K.}
\newblock \bibinfo{title}{Mixed-state entanglement and quantum error
  correction}.
\newblock \emph{\bibinfo{journal}{Phys. Rev. A}} \textbf{\bibinfo{volume}{54}},
  \bibinfo{pages}{3824--3851} (\bibinfo{year}{1996}).

\bibitem{horodecki1998mixed}
\bibinfo{author}{Horodecki, M.}, \bibinfo{author}{Horodecki, P.} \&
  \bibinfo{author}{Horodecki, R.}
\newblock \bibinfo{title}{{M}ixed-{S}tate {E}ntanglement and {D}istillation:
  {I}s there a {``Bound''} {E}ntanglement in {N}ature?}
\newblock \emph{\bibinfo{journal}{Phys. Rev. Lett.}}
  \textbf{\bibinfo{volume}{80}}, \bibinfo{pages}{5239--5242}
  (\bibinfo{year}{1998}).

\bibitem{Horodecki2005a}
\bibinfo{author}{Horodecki, K.}, \bibinfo{author}{Horodecki, M.},
  \bibinfo{author}{Horodecki, P.} \& \bibinfo{author}{Oppenheim, J.}
\newblock \bibinfo{title}{{S}ecure {K}ey from {B}ound {E}ntanglement}.
\newblock \emph{\bibinfo{journal}{Phys. Rev. Lett.}}
  \textbf{\bibinfo{volume}{94}}, \bibinfo{pages}{160502}
  (\bibinfo{year}{2005}).

\bibitem{Horodecki2008a}
\bibinfo{author}{Horodecki, K.}, \bibinfo{author}{Pankowski, {\L}.},
  \bibinfo{author}{Horodecki, M.} \& \bibinfo{author}{Horodecki, P.}
\newblock \bibinfo{title}{{L}ow-{D}imensional {B}ound {E}ntanglement {W}ith
  {O}ne-{W}ay {D}istillable {C}ryptography {K}ey}.
\newblock In \emph{\bibinfo{booktitle}{{IEEE} {T}ransactions on {I}nformation
  {T}heory}}, vol.~\bibinfo{volume}{54}, \bibinfo{pages}{2621--2625}
  (\bibinfo{year}{2008}).

\bibitem{Smith2008a}
\bibinfo{author}{Smith, G.} \& \bibinfo{author}{Yard, J.}
\newblock \bibinfo{title}{{Q}uantum {C}ommunication with {Z}ero-{C}apacity
  {C}hannels}.
\newblock \emph{\bibinfo{journal}{Science}} \textbf{\bibinfo{volume}{321}},
  \bibinfo{pages}{1812--1815} (\bibinfo{year}{2008}).

\bibitem{amselem2009experimental}
\bibinfo{author}{Amselem, E.} \& \bibinfo{author}{Bourennane, M.}
\newblock \bibinfo{title}{{E}xperimental four-qubit bound entanglement}.
\newblock \emph{\bibinfo{journal}{Nat. Phys.}} \textbf{\bibinfo{volume}{5}},
  \bibinfo{pages}{748--752} (\bibinfo{year}{2009}).

\bibitem{smolin2001four}
\bibinfo{author}{Smolin, J.~A.}
\newblock \bibinfo{title}{{F}our-party unlockable bound entangled state}.
\newblock \emph{\bibinfo{journal}{Phys. Rev. A}} \textbf{\bibinfo{volume}{63}},
  \bibinfo{pages}{032306} (\bibinfo{year}{2001}).

\bibitem{yang2005irreversibility}
\bibinfo{author}{Yang, D.}, \bibinfo{author}{Horodecki, M.},
  \bibinfo{author}{Horodecki, R.} \& \bibinfo{author}{Synak-Radtke, B.}
\newblock \bibinfo{title}{{I}rreversibility for {A}ll {B}ound {E}ntangled
  {S}tates}.
\newblock \emph{\bibinfo{journal}{Phys. Rev. Lett.}}
  \textbf{\bibinfo{volume}{95}}, \bibinfo{pages}{190501}
  (\bibinfo{year}{2005}).

\bibitem{piani2009relative}
\bibinfo{author}{Piani, M.}
\newblock \bibinfo{title}{{R}elative {E}ntropy of {E}ntanglement and
  {R}estricted {M}easurements}.
\newblock \emph{\bibinfo{journal}{Phys. Rev. Lett.}}
  \textbf{\bibinfo{volume}{103}}, \bibinfo{pages}{160504}
  (\bibinfo{year}{2009}).

\bibitem{brandao2010generalization}
\bibinfo{author}{Brand{\~{a}}o, F. G. S.~L.} \& \bibinfo{author}{Plenio, M.}
\newblock \bibinfo{title}{{A} {G}eneralization of {Q}uantum {{S}tein's}
  {L}emma}.
\newblock \emph{\bibinfo{journal}{{C}ommunications in {M}athematical
  {P}hysics}} \textbf{\bibinfo{volume}{295}}, \bibinfo{pages}{791--828}
  (\bibinfo{year}{2010}).

\bibitem{horodecki1997separability}
\bibinfo{author}{Horodecki, P.}
\newblock \bibinfo{title}{{S}eparability criterion and inseparable mixed states
  with positive partial transposition}.
\newblock \emph{\bibinfo{journal}{Physics Letters A}}
  \textbf{\bibinfo{volume}{232}}, \bibinfo{pages}{333--339}
  (\bibinfo{year}{1997}).

\bibitem{masanes2006all}
\bibinfo{author}{Masanes, L.}
\newblock \bibinfo{title}{{A}ll {B}ipartite {E}ntangled {S}tates {A}re {U}seful
  for {I}nformation {P}rocessing}.
\newblock \emph{\bibinfo{journal}{Phys. Rev. Lett.}}
  \textbf{\bibinfo{volume}{96}}, \bibinfo{pages}{150501}
  (\bibinfo{year}{2006}).

\bibitem{piani2009all}
\bibinfo{author}{Piani, M.} \& \bibinfo{author}{Watrous, J.}
\newblock \bibinfo{title}{{A}ll {E}ntangled {S}tates are {U}seful for {C}hannel
  {D}iscrimination}.
\newblock \emph{\bibinfo{journal}{Phys. Rev. Lett.}}
  \textbf{\bibinfo{volume}{102}}, \bibinfo{pages}{250501}
  (\bibinfo{year}{2009}).

\bibitem{augusiak2006bound}
\bibinfo{author}{Augusiak, R.} \& \bibinfo{author}{Horodecki, P.}
\newblock \bibinfo{title}{{B}ound entanglement maximally violating {B}ell
  inequalities: {Q}uantum entanglement is not fully equivalent to cryptographic
  security}.
\newblock \emph{\bibinfo{journal}{Phys. Rev. A}} \textbf{\bibinfo{volume}{74}},
  \bibinfo{pages}{010305} (\bibinfo{year}{2006}).

\bibitem{Kampermann09nmr}
\bibinfo{author}{Kampermann, H.}, \bibinfo{author}{Bru\ss{}, D.},
  \bibinfo{author}{Peng, X.} \& \bibinfo{author}{Suter, D.}
\newblock \bibinfo{title}{Experimental generation of
  pseudo-bound-entanglement}.
\newblock \emph{\bibinfo{journal}{Phys. Rev. A}} \textbf{\bibinfo{volume}{81}},
  \bibinfo{pages}{040304(R)} (\bibinfo{year}{2010}).

\bibitem{Hong1987a}
\bibinfo{author}{Hong, C.~K.}, \bibinfo{author}{Ou, Z.~Y.} \&
  \bibinfo{author}{Mandel, L.}
\newblock \bibinfo{title}{Measurement of subpicosecond time intervals between
  two photons by interference}.
\newblock \emph{\bibinfo{journal}{Phys. Rev. Lett.}}
  \textbf{\bibinfo{volume}{59}}, \bibinfo{pages}{2044--2046}
  (\bibinfo{year}{1987}).

\bibitem{Jezek2003a}
\bibinfo{author}{Je{\v{z}}ek, M.}, \bibinfo{author}{Fiur{\'a}{\v{s}}ek, J.} \&
  \bibinfo{author}{Hradil, Z.}
\newblock \bibinfo{title}{Quantum inference of states and processes}.
\newblock \emph{\bibinfo{journal}{Phys. Rev. A}} \textbf{\bibinfo{volume}{68}},
  \bibinfo{pages}{012305} (\bibinfo{year}{2003}).

\bibitem{Jozsa1994a}
\bibinfo{author}{Jozsa, R.}
\newblock \bibinfo{title}{{F}idelity for {M}ixed {Q}uantum {S}tates}.
\newblock \emph{\bibinfo{journal}{J. Mod. Opt.}} \textbf{\bibinfo{volume}{41}},
  \bibinfo{pages}{2315--2323} (\bibinfo{year}{1994}).

\bibitem{Werner89}
\bibinfo{author}{Werner, R.~F.}
\newblock \bibinfo{title}{Quantum states with einstein-podolsky-rosen
  correlations admitting a hidden-variable model}.
\newblock \emph{\bibinfo{journal}{Phys. Rev. A}} \textbf{\bibinfo{volume}{40}},
  \bibinfo{pages}{4277--4281} (\bibinfo{year}{1989}).

\bibitem{Kwiat1999a}
\bibinfo{author}{Kwiat, P.~G.}, \bibinfo{author}{Waks, E.},
  \bibinfo{author}{White, A.~G.}, \bibinfo{author}{Appelbaum, I.} \&
  \bibinfo{author}{Eberhard, P.~H.}
\newblock \bibinfo{title}{{U}ltrabright source of polarization-entangled
  photons}.
\newblock \emph{\bibinfo{journal}{Phys. Rev. A}} \textbf{\bibinfo{volume}{60}},
  \bibinfo{pages}{R773--R776} (\bibinfo{year}{1999}).

\bibitem{Lavoie2009a}
\bibinfo{author}{Lavoie, J.}, \bibinfo{author}{Kaltenbaek, R.} \&
  \bibinfo{author}{Resch, K.~J.}
\newblock \bibinfo{title}{{E}xperimental violation of {S}vetlichny's
  inequality}.
\newblock \emph{\bibinfo{journal}{New J. Phys.}} \textbf{\bibinfo{volume}{11}},
  \bibinfo{pages}{073051} (\bibinfo{year}{2009}).

\bibitem{pittinger_rubin2002}
\bibinfo{author}{Pittenger, A.~O.} \& \bibinfo{author}{Rubin, M.~H.}
\newblock \bibinfo{title}{{C}onvexity and the separability problem of quantum
  mechanical density matrices}.
\newblock \emph{\bibinfo{journal}{Linear Algebra and its Applications}}
  \textbf{\bibinfo{volume}{346}}, \bibinfo{pages}{47--71}
  (\bibinfo{year}{2002}).

\bibitem{Doherty2009a}
\bibinfo{author}{Doherty, A.~C.}, \bibinfo{author}{Gilchrist, A.} \&
  \bibinfo{author}{de~Burgh, M.~D.} (\bibinfo{year}{2009}).
\newblock \bibinfo{note}{Unpublished}.

\end{thebibliography}
\end{document}